\documentclass[a4paper,10pt]{elsarticle}

\usepackage[margin=2.25cm]{geometry}
\usepackage{natbib}
\usepackage{floatrow}   
\usepackage{graphicx,xcolor} 
\usepackage{dcolumn}
\usepackage{bm}
\usepackage[hidelinks]{hyperref}
\usepackage{cancel}
\usepackage{caption}
\usepackage{subfigure}
\usepackage{commath}
\usepackage{amsmath}
\usepackage{verbatim}
\usepackage{color}
\usepackage{bigints}
\usepackage{lineno}

\DeclareMathOperator{\erf}{erf}

\makeatletter
\def\@author#1{\g@addto@macro\elsauthors{\normalsize
    \def\baselinestretch{1}
    \upshape\authorsep#1\unskip\textsuperscript{
      \ifx\@fnmark\@empty\else\unskip\sep\@fnmark\let\sep=,\fi
      \ifx\@corref\@empty\else\unskip\sep\@corref\let\sep=,\fi
      }
    \def\authorsep{\unskip,\space}
    \global\let\@fnmark\@empty
    \global\let\@corref\@empty
    \global\let\sep\@empty}
    \@eadauthor={#1}
}
\makeatother

\begin{document}

\begin{frontmatter}

\title{Extrapolation Technique Pitfalls in Asymmetry Measurements at Colliders}

\author{Katrina Colletti\corref{cor1}}
\ead{kcolletti1@tamu.edu}
\cortext[cor1]{Corresponding author}

\author{Ziqing Hong}

\author{David Toback}

\author{Jonathan S. Wilson}

\address{The George P. and Cynthia Woods Mitchell Institute for Fundamental Physics and Astronomy, Texas A\&M University\\
College Station, TX 77843-4242}

\begin{abstract}
Asymmetry measurements are common in collider experiments and can sensitively probe particle properties. Typically, data can only be measured in a finite region covered by the detector, so an extrapolation from the visible asymmetry to the inclusive asymmetry is necessary. Often a constant multiplicative factor is more than adequate for the extrapolation and this factor can be readily determined using simulation methods. However, there is a potential, avoidable pitfall involved in the determination of this factor when the asymmetry in the simulated data sample is small. We find that to obtain a reliable estimate of the extrapolation factor, the number of simulated events required rises as the inverse square of the simulated asymmetry; this can mean that an unexpectedly large sample size is required when determining its value.
\end{abstract}

\begin{keyword}
asymmetry \sep linear extrapolation \sep Monte Carlo \sep collider experiments
\end{keyword}

\end{frontmatter}

\section{\label{sec:intro}Introduction}
Measurements of production asymmetries have a long history at colliders, so examination of some of the experimental techniques used to make them is important. Most measurements are performed by first measuring the asymmetry within a restricted geometric region -- the region covered by the detector -- and then extrapolating to the inclusive region. In some cases a constant multiplicative factor can reliably be used. Because this sort of technique is widely applicable to experimental measurements, we explore it in detail here and identify an important potential pitfall in estimating the multiplicative factor via simulations.

In general an asymmetry is defined with the partial cross sections, \(\sigma_{1}\) and \(\sigma_{2}\), over two complementary kinematic or geometric regions,
\begin{linenomath}
\begin{align}
A&\equiv\frac{\sigma_{1}-\sigma_{2}}{\sigma_{1}+\sigma_{2}}\label{eqn:first}\,.
\end{align}
\end{linenomath}
We can simplify our discussion by considering the regions defined by a single variable, \(x\), while integrating over all other variables. 
In the case where \(x\) represents the pseudorapidity of a particle, which is directly related to the angle \(\theta\) between an outgoing particle and the beam line, this produces a forward-backward asymmetry, for example for use in top-quark-pair production at the Fermilab Tevatron~\cite{costheta,CDFresult,d0Measurement2,DZEROResult,gausApprox}. We define \(A^{\text{inclusive}}\) using
\begin{linenomath}
\begin{align}
\sigma_{1}^{\text{inclusive}}&=\int_{0}^{\infty}dx\,\frac{d\sigma}{dx},\text{ and }\nonumber\\
\sigma_{2}^{\text{inclusive}}&=\int_{-\infty}^{0}dx\,\frac{d\sigma}{dx}.
\end{align}
\end{linenomath}
However, when the entire range of \(x\) is not accessible due to kinematic constraints and/or the geometry of the detector, we can only measure
\begin{linenomath}
\begin{align}
\sigma_{1}^{\text{visible}}&=\int_0^{x^{\text{visible}}}dx\,\frac{d\sigma}{dx},\text{ and }\nonumber\\
\sigma_{2}^{\text{visible}}&=\int_{-x^{\text{visible}}}^{0}dx\,\frac{d\sigma}{dx},
\end{align}
\end{linenomath}
which define the visible asymmetry, \(A^{\text{visible}}\).

There are multiple ways to extrapolate from \(A^{\text{visible}}\) to \(A^{\text{inclusive}}\). The two simplest methods for doing this are employing an additive correction factor (C=\(A^{\text{inclusive}}-A^{\text{visible}}\))~\cite{cdfWillisAdditive,d0willisversion} or, a method that is commonly used, employing a multiplicative correction factor
\begin{linenomath}
\begin{align}
R&=\frac{A^{\text{visible}}}{A^{\text{inclusive}}},\label{eqn:Rdef}
\end{align}
\end{linenomath}
where each are typically estimated using Monte Carlo (MC) simulations~\cite{CDFresult,d0Measurement2,ziqing2014}. Each is applicable in different physical scenarios. While more sophisticated correction methods can be employed~\cite{costheta,ATLASAfb,ATLASnote,ATLAS2,ATLAS3,ATLAS4,CMSAfb,CMSsoph,CMS2014jua,LHCbBBAC,CDFLJAfbl,CDFLJAfb,D0BmesonAfb,bbbar,DZEROResult2}, the multiplicative correction method has been very successful for \(t\bar{t}\) leptonic asymmetry measurements, as the correction factor appears not to vary significantly with the inclusive asymmetry~\cite{ziqing2014}. In this Article, we explore a simple example in which this condition holds, but use it to identify a pitfall in the estimation of the correction factor and explore ways in which this pitfall may be avoided by future analyses.

For illustrative purposes, we consider a simplified model based on the measurement of the top leptonic forward-backward asymmetry at the Fermilab Tevatron~\cite{CDFresult,d0Measurement2,DZEROResult,gausApprox}. It has been shown both that the differential cross section of leptons as a function of pseudorapidity can be well approximated as the sum of two Gaussian distributions with a common mean, and that the simple multiplicative extrapolation technique works in this case~\cite{ziqing2014}. For the purposes of this study, we take the differential cross section \(\frac{d\sigma}{dx}\) to be the simpler single-Gaussian distribution with unit width and a non-zero mean, \(\mu\). As shown in \ref{sec:closedNumSoln} there is an approximately linear relationship between the asymmetry and \(\mu\) for small values of \(\mu\); we can refer to the behavior of \(\mu\) and the asymmetry interchangeably. This simple model provides a foundation to understand the general behavior of multiplicative asymmetry extrapolation methods.

\begin{figure}[htb]
\subfigure[]{
	\includegraphics[width=0.47\columnwidth, keepaspectratio]{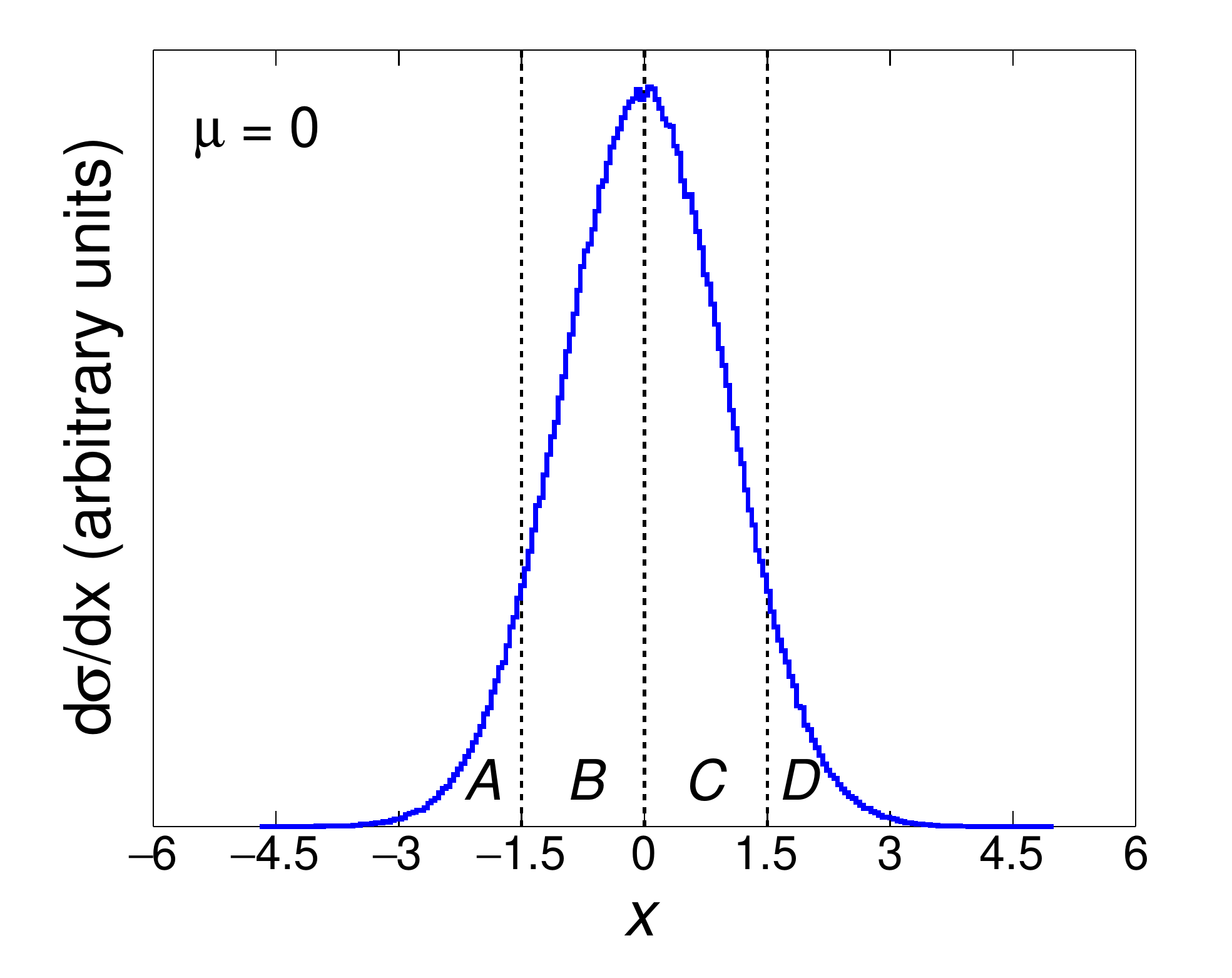}
	\label{subfig:gausMu0Ev1e6}
}
\subfigure[]{
	\includegraphics[width=0.47\columnwidth, keepaspectratio]{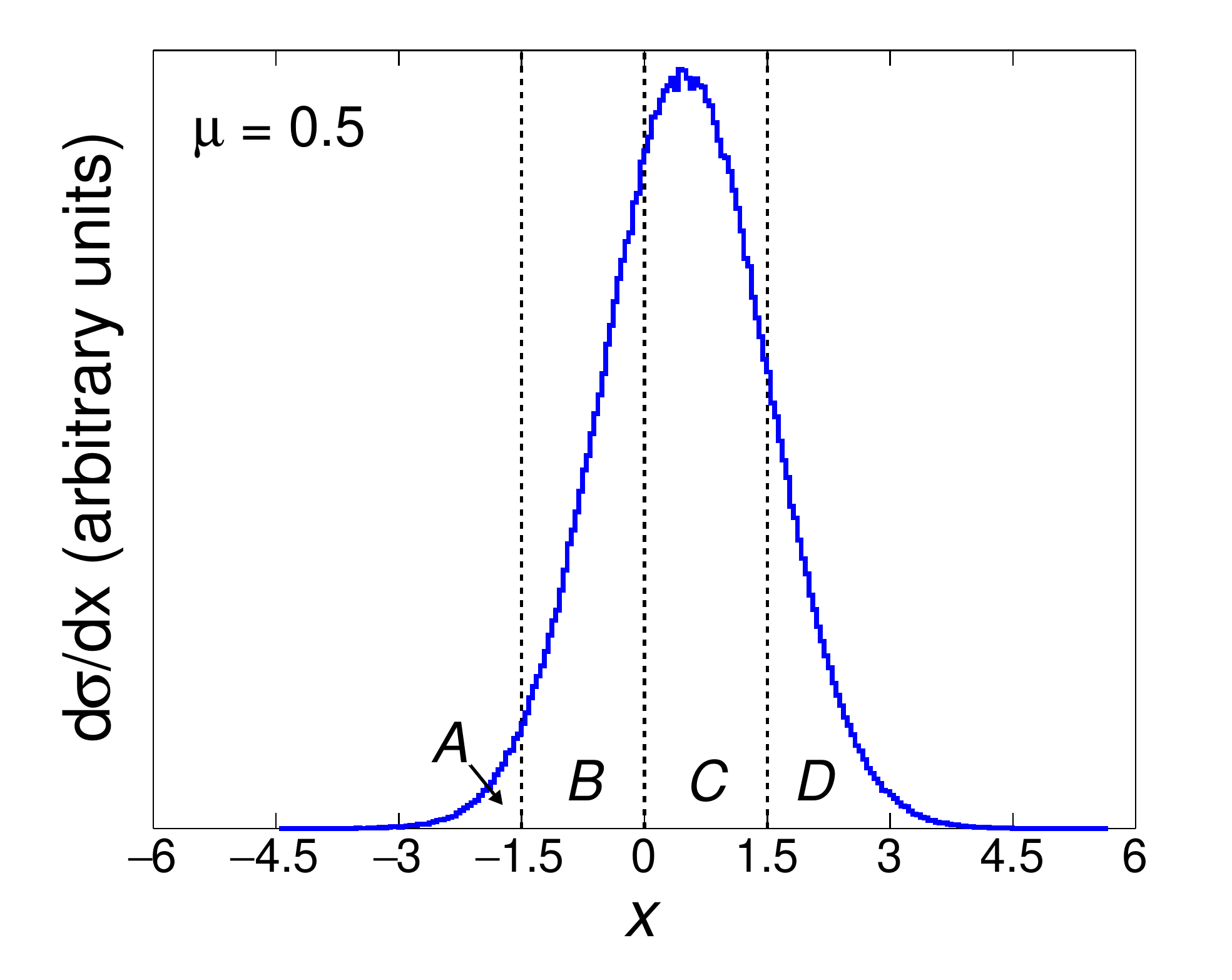}
	\label{subfig:gausMuPt5Ev1e6}
}
\caption{Two Gaussian distributions with unit width, with \(\mu=0.0\) and \(\mu=0.5\) in (a) and (b) respectively. We highlight the events in regions \(A\,(-\infty,-1.5),\,B\,(-1.5,0),\, C\,(0,1.5),\text{ and }D\,(1.5,\infty)\).}
\label{fig:simpleGaus}
\end{figure}

A potential pitfall occurs when estimating the correction factor in Eq.~(\ref{eqn:Rdef}) using MC samples with small asymmetries. Under certain quantifiable conditions, simulations can produce values of \(R\) that are misleading and far from the correct value. To make the discussion concrete, we pick a visible region for our single Gaussian distribution of \(-1.5<x<1.5\), which gives the visible and inclusive regions as shown in Fig.~\ref{fig:simpleGaus}. Given this particular description, to an excellent degree of approximation we find \(R=0.7795\pm0.0005\), as shown in \ref{sec:closedNumSoln}. Since analyses typically have more complicated distributions and use MC methods to estimate \(R\), we begin this study by using MC samples to determine the distribution of the multiplicative factor, and illustrate the pitfalls when the simulated \(A^{\text{inclusive}}\) goes to zero. We then compare this result with a closed form statistical solution to gain a better understanding of why this pitfall arises.

\section{\label{sec:MCsoln}Monte Carlo Study}
\begin{figure}[htbp]
\subfigure[]{
\centering
	\includegraphics[width=0.47\columnwidth, keepaspectratio]{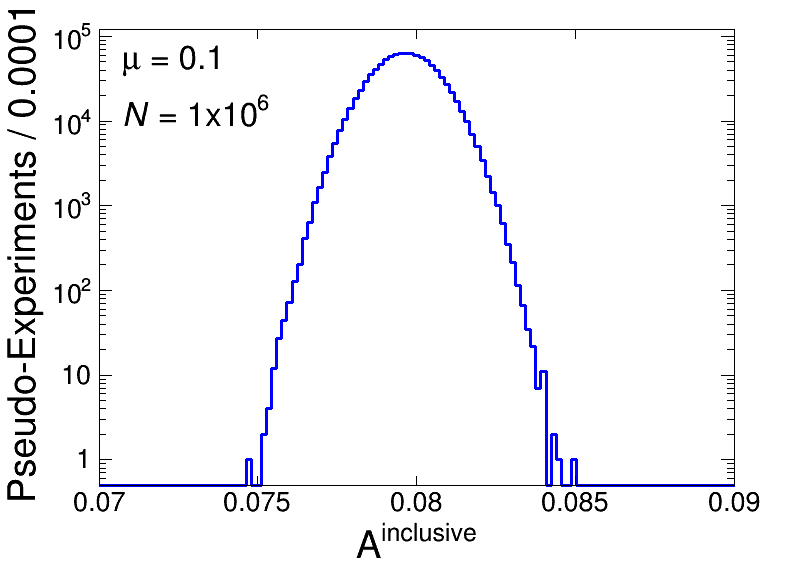}
	\label{subfig:tot_muPt1_ev1e+6}
}
\subfigure[]{
\centering
	\includegraphics[width=0.47\columnwidth, keepaspectratio]{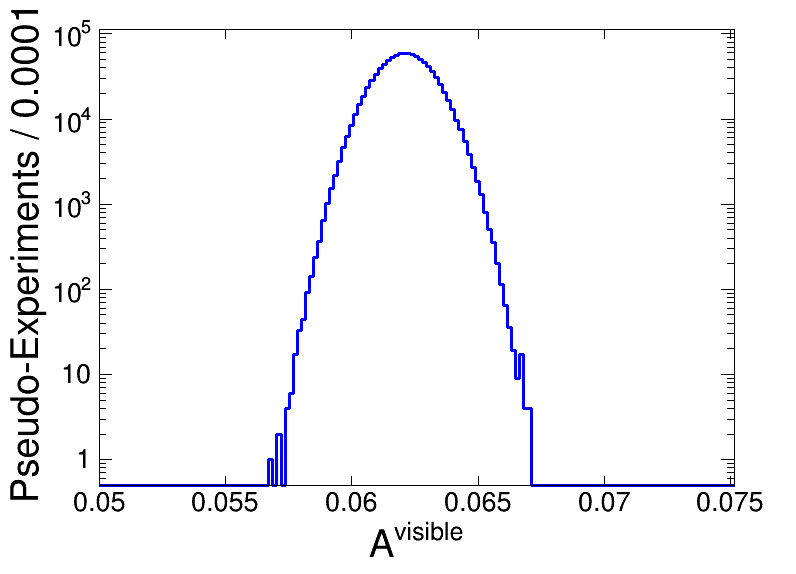}
	\label{subfig:red_muPt1_ev1e+6}
}
\\[2ex]
\subfigure[]{
\centering
	\includegraphics[width=0.47\columnwidth, keepaspectratio]{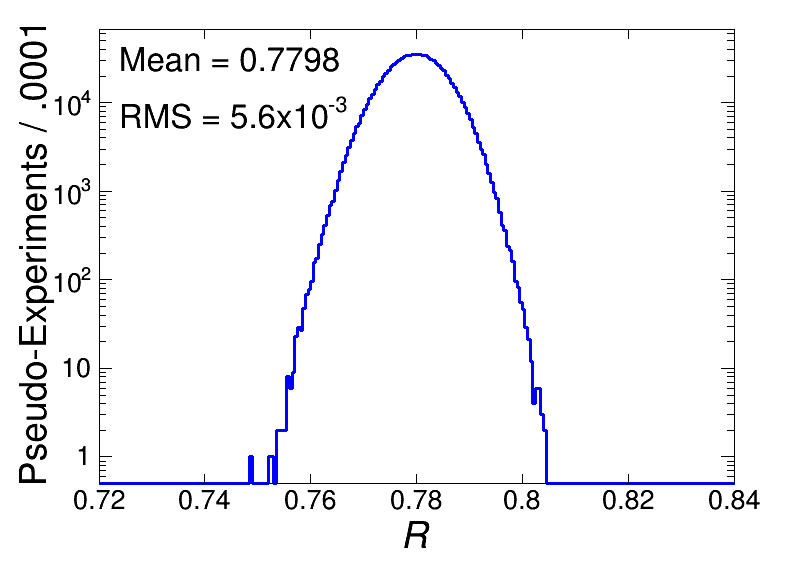}
	\label{subfig:ratio_muPt1_ev1e+6}
}
\caption{Distributions of \(A^{\text{inclusive}}\), \(A^{\text{visible}}\), and \(R\) in (a), (b), and (c) respectively. Each distribution has \(N_{\text{PE}}=10^{6}\) with \(N=10^{6}\) and \(\mu=0.1\).}
\label{fig:tot_red_ratio_muPt1_ev1e+6}
\end{figure}

The most common method to determine the multiplicative correction factor is to simulate events according to a calculated differential cross section \(\frac{d\sigma}{dx}\), and calculate the correction factor \(R\) from the simulated events. We mimic this procedure by generating sets of random numbers according to a simplified differential cross section that takes the form of a Gaussian function with unit width and a mean \(\mu\). Each random number represents an event, each set of random numbers is a pseudo-experiment (PE), and the number of events in each PE is denoted by \(N\). From each PE, we can measure both \(A^{\text{visible}}\) and \(A^{\text{inclusive}}\), and therefore \(R\). Distributions of these three values can then be generated with an ensemble of PEs; the number of PEs used to generate these distributions is denoted by \(N_{\text{PE}}\). For example, in Fig.~\ref{subfig:gausMu0Ev1e6}, we show our differential cross section, a single PE, with \(N=10^{6}\) and \(\mu=0\). In Fig.~\ref{fig:tot_red_ratio_muPt1_ev1e+6}, we show the distributions of \(A^{\text{visible}}\), \(A^{\text{inclusive}}\), and \(R\) for \(N_{\text{PE}}=10^{6}\), each with \(N=10^{6}\) and \(\mu=0.1\). This value of \(\mu\) is chosen as it corresponds to \(A^{\text{inclusive}}\approx8\%\), which is a value we typically see in \(t\bar{t}\) asymmetry measurements at the Tevatron~\cite{CDFresult,d0Measurement2,DZEROResult,gausApprox}. 

Since the simulation of a practical differential cross section is usually computationally expensive, the common practice is to simulate one PE with a modest \(N\), usually on the order of \(10^{6}\), and calculate \(R\) from it. In this analysis, the distribution of \(R\) from an ensemble of PEs reveals the quality of the estimation of \(R\) from a single PE. We note that in Fig.~\ref{subfig:ratio_muPt1_ev1e+6} the variation in \(R\) is small, with a width less than \(1\%\) of its mean value. With the simplified single-Gaussian differential cross section and the visible region specified above, \(R=0.7798\) which is consistent with the calculation in~\ref{sec:closedNumSoln}.

We next study the quality of the estimation of \(R\) as we vary the two factors, \(\mu\) and \(N\), which have significant impact on potential measurements: we examine what happens both in the limit of small simulation sample size and as \(\mu\rightarrow0\) (or equivalently, as the asymmetry approaches zero). Specifically, we aim to understand whether the estimation of \(R\) is correct and what the uncertainty on that estimation is, the sample size needed to obtain a small uncertainty, and whether the value of \(R\) is constant for all values of \(\mu\) when it is measured with a large sample size.

\begin{figure}[htbp]
\subfigure[]{
\centering
	\includegraphics[width=0.47\columnwidth, keepaspectratio]{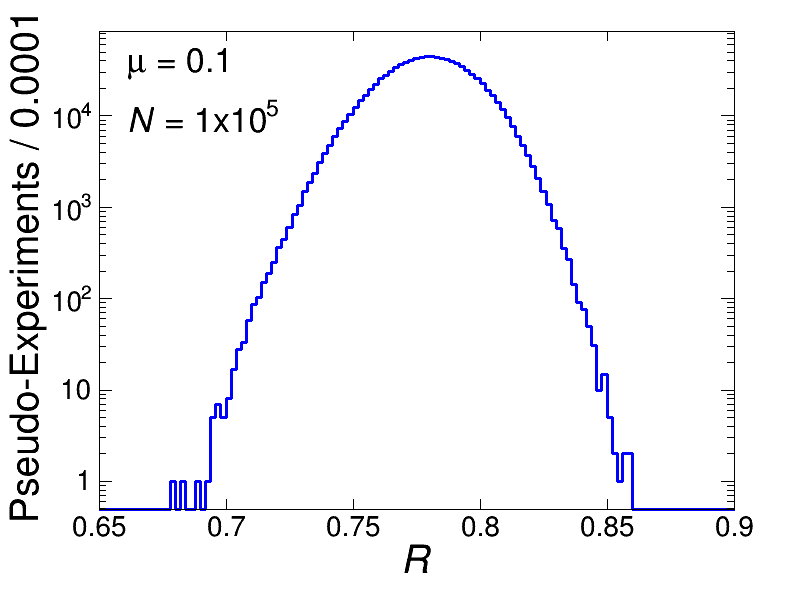}
	\label{subfig:ratio_muPt1_ev1e+5}
}
\subfigure[]{
\centering
	\includegraphics[width=0.47\columnwidth, keepaspectratio]{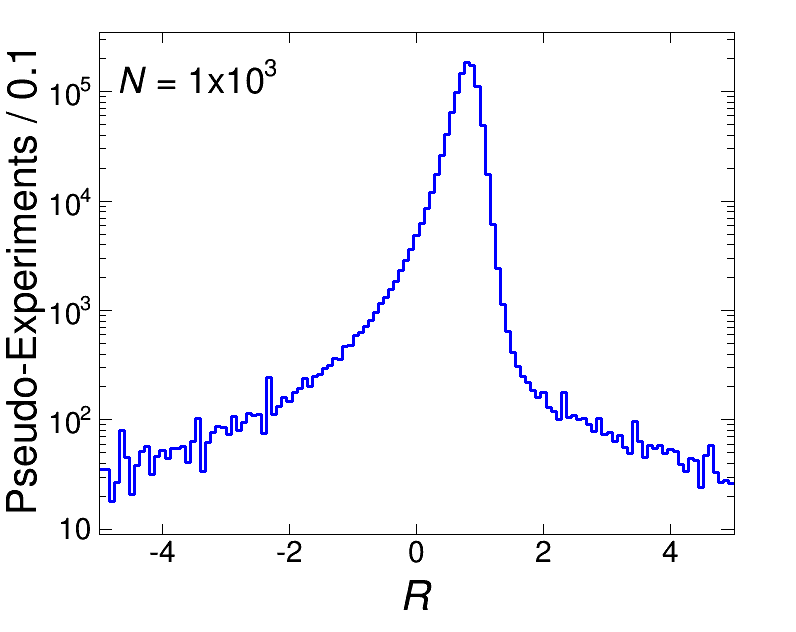}
	\label{subfig:ratio_muPt1_ev1e+3}
}
\caption{Distributions of \(R\) with \(N_{\text{PE}}=10^{6}\) and \(\mu=0.1\), for \(N=10^{5}\) and \(N=10^{3}\) in (a) and (b) respectively. As \(N\) decreases, the estimation of \(R\) becomes worse; therefore, obtaining the correct result with a single PE becomes statistically unreliable, and the systematic uncertainty becomes both significantly large and asymmetric. Note the different x-axis scales in both plots.}
\label{fig:ratio_muPt1_varyingEvents}
\end{figure}

For \(\mu=0.1\), \(R\) is well determined even with a fairly small value of \(N\). Figure~\ref{fig:ratio_muPt1_varyingEvents} shows distributions of \(R\) for \(N_{\text{PE}}=10^{6}\) with \(N=10^{5}\) and \(N=10^{3}\). As \(N\) decreases, the \(R\) distribution becomes much wider and less Gaussian, and estimating the value of \(R\) from a single PE (as is typically done in realistic scenarios with more complicated differential cross sections) quickly leads to incorrect results. Note that the peak of the distribution still appears at the same place for reasons that will be discussed in Sec.~\ref{sec:closedStatSoln}. Thus, it becomes clear that there is a minimum allowable \(N\), above which we can be confident in the estimation of \(R\), and below which the estimation of \(R\) is no longer reliable, thus introducing a significant systematic uncertainty to the inclusive asymmetry measurement.

\begin{figure}[htbp]
\subfigure[]{
\centering
	\includegraphics[width=0.47\columnwidth, keepaspectratio]{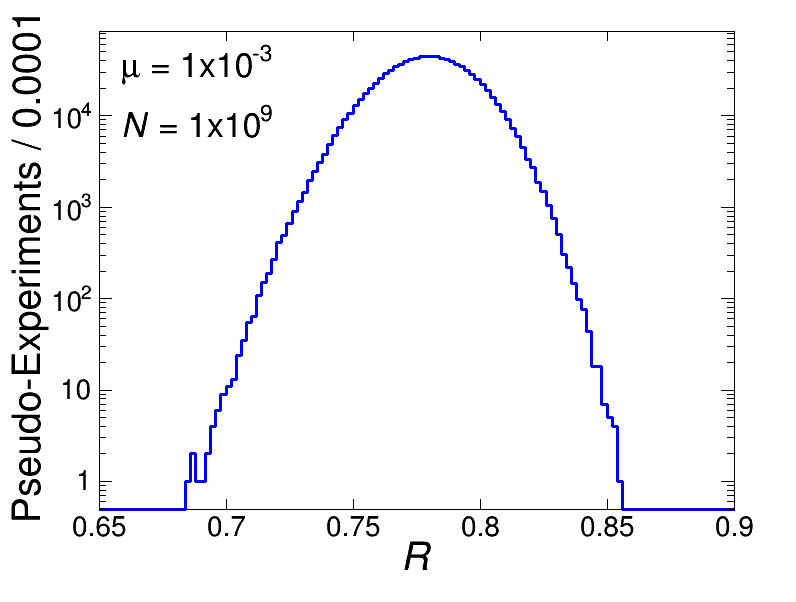}
	\label{subfig:ratio_muPt001_ev1e+9}
}
\subfigure[]{
\centering
	\includegraphics[width=0.47\columnwidth, keepaspectratio]{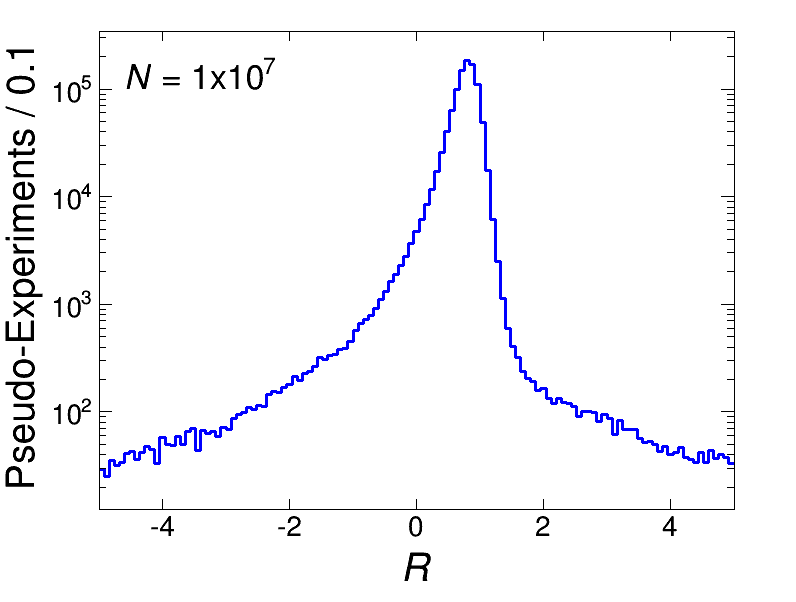}
	\label{subfig:ratio_muPt001_ev1e+7}
}
\caption{The same set of plots as in Fig.~\ref{fig:ratio_muPt1_varyingEvents}, but for \(\mu=10^{-3}\) with \(N=10^{9}\) and \(N=10^{7}\) in (a) and (b) respectively. We note that the distribution transition also occurs for this \(\mu\), but at a larger value of \(N\).}
\label{fig:ratio_muPt01_varyingEvents}
\end{figure}

This issue becomes even more pronounced as \(\mu\), and thus the asymmetry, approaches 0. In Fig.~\ref{fig:ratio_muPt01_varyingEvents}, we show distributions of the estimated value of \(R\) but for \(\mu=10^{-3}\) and larger values of \(N_{\text{PE}}\). The first thing we note is that \(R\) is virtually identical when the simulation size is sufficient. However, we also notice that it requires four orders of magnitude more events to get the same width in the distribution of \(R\) as it did for \(\mu=0.1\). As we note in the next section, this effect has been studied in great detail in the statistics literature, and we see that the \(R\) distribution begins to approximate a Cauchy distribution \cite{Cauchy}. The usual measurements of mean and standard deviation are not expected to give accurate and reliable results; indeed, for a true Cauchy distribution these two values are not defined.

\begin{figure}[htbp]
	\includegraphics[width=0.47\columnwidth, keepaspectratio]{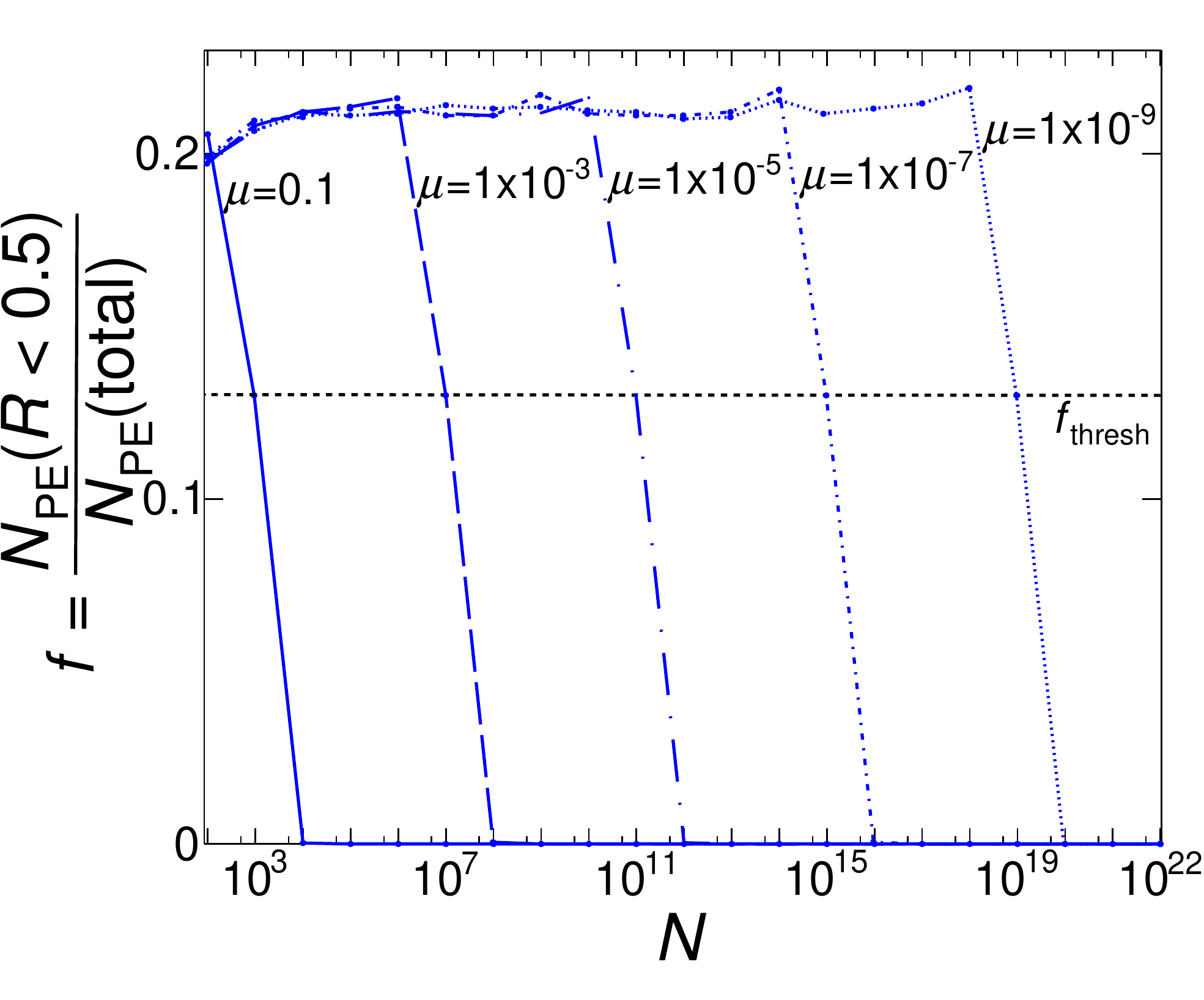}
	\label{subfig:fracMu1e-8}
\caption{A plot showing \(f\), the fraction of PEs with \(R<0.5\), versus \(N\). Each line represents a different choice of \(\mu\) varying from \(\mu=0.1\) to \(\mu=10^{-9}\). We highlight \(f_{\text{thresh}}=0.13\), and note that as \(\mu\) gets smaller, the value of \(N\) where the line crosses \(f_{\text{thresh}}\) gets significantly larger.}
\label{fig:frac_plots}
\end{figure}

To determine how many events we need to be able to make a reliable estimation of \(R\), we define the fraction of PEs with \(R<0.5\):
\begin{equation}
f=\frac{N_\text{PE}(R<0.5)}{N_\text{PE}(\text{total})}.
\end{equation}
The choice here of \(R<0.5\) is somewhat arbitrary, and the results do not depend on this choice. However, it is chosen to capture information on the lower tail of the \(R\) distribution and since it is many standard deviations away from the large \(N\) answer we require \(f\approx0\) for a reliable measurement of \(R\). In Fig.~\ref{fig:frac_plots} we show \(f\) varying with \(N\) for various values of \(\mu\). For each value of \(\mu\), we see the same basic structure. For low statistics (small \(N\)) we see large values of \(f\) (typically above 20\%). However, at some threshold, \(f\) drops quickly to zero. A qualitative definition of ``proper statistics'' is requiring \(N\) to be in the region where \(f\sim0\), which we refer to as the ``high-statistics regime''; otherwise we are in the ``low-statistics regime''.

\begin{figure}[htbp]
    \begin{center}
        \includegraphics[width=0.47\textwidth, keepaspectratio]{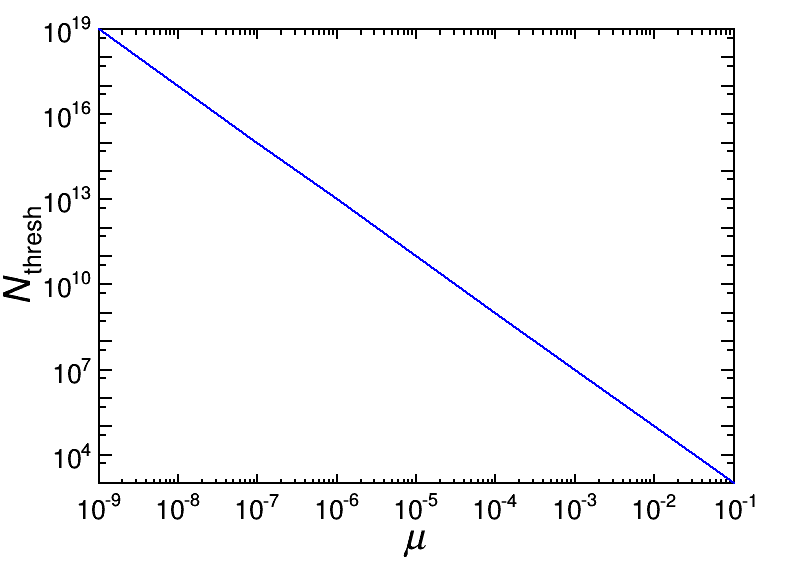}
\caption{A plot of \(N_{\text{thresh}}\) versus \(\mu\). Note that as \(\mu\rightarrow0\), \(N_{\text{thresh}}\rightarrow\infty\).}
\label{fig:threshold}
\end{center}
\end{figure}

For \(\mu=0.1\), \(f\) approaches 0 at \(N\sim10^{3}\). This is consistent with what we see in Fig.~\ref{fig:ratio_muPt1_varyingEvents}; it is Gaussian for \(N=10^{5}\), but begins approximating a Cauchy distribution at \(N\sim10^{3}\). For \(\mu=10^{-3}\), \(f\) goes to 0 at \(N\sim10^{7}\) (as seen in Fig.~\ref{fig:ratio_muPt01_varyingEvents}), and so on. This indicates that as \(\mu\) approaches zero, the number of events that are needed to reliably estimate \(R\) increases. We quantify a measure of this ``threshold'' by defining \(f_{\text{thresh}}=0.13\) and then measuring the corresponding \(N\), which we denote as \(N_{\text{thresh}}\). This choice of \(f_\text{thresh}\) is also somewhat arbitrary, as any value between \(0\%\) and \(\approx20\%\) would capture the relationship between \(N_\text{thresh}\) and \(\mu\) that we seek, and thus the results do not depend on this choice. We examine how this value varies as \(\mu\rightarrow0\). The result is shown in Fig.~\ref{fig:threshold}. As we can read off from the graph (and will show analytically in Sec.~\ref{sec:closedStatSoln}), \(\mu\) is roughly proportional to \(1/\sqrt{N_{\text{thresh}}}\). We note that \(N_{\text{thresh}}\rightarrow\infty\) as \(\mu\rightarrow0\).

\begin{figure}[htbp]
\subfigure[]{
\centering
	\includegraphics[width=0.47\columnwidth, keepaspectratio]{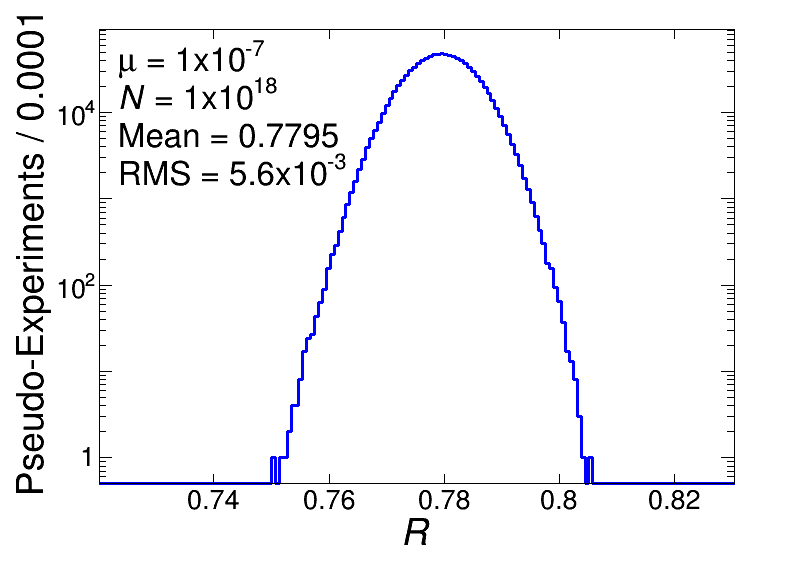}
	\label{subfig:ratio_mu1e-7_ev1e+18}
}
\subfigure[]{
\centering
	\includegraphics[width=0.47\columnwidth, keepaspectratio]{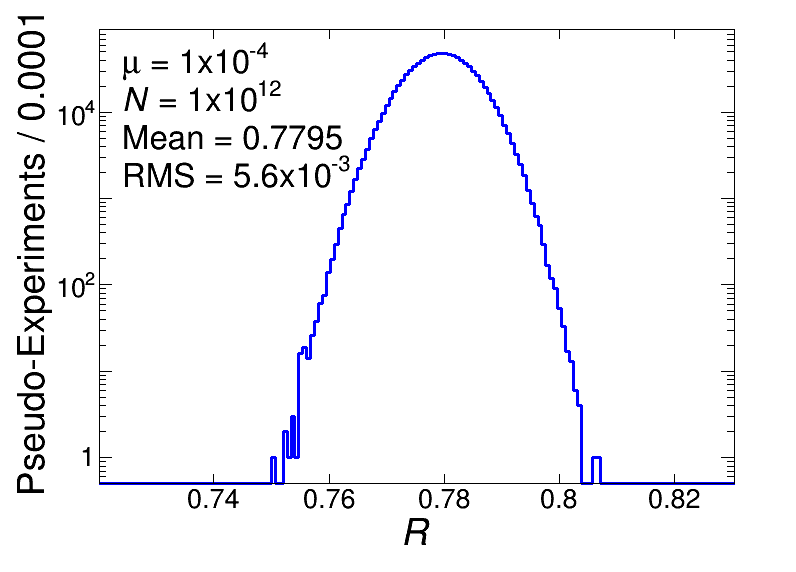}
	\label{subfig:ratio_muPtE-4_ev1e+12}
}
\subfigure[]{
\centering
	\includegraphics[width=0.47\columnwidth, keepaspectratio]{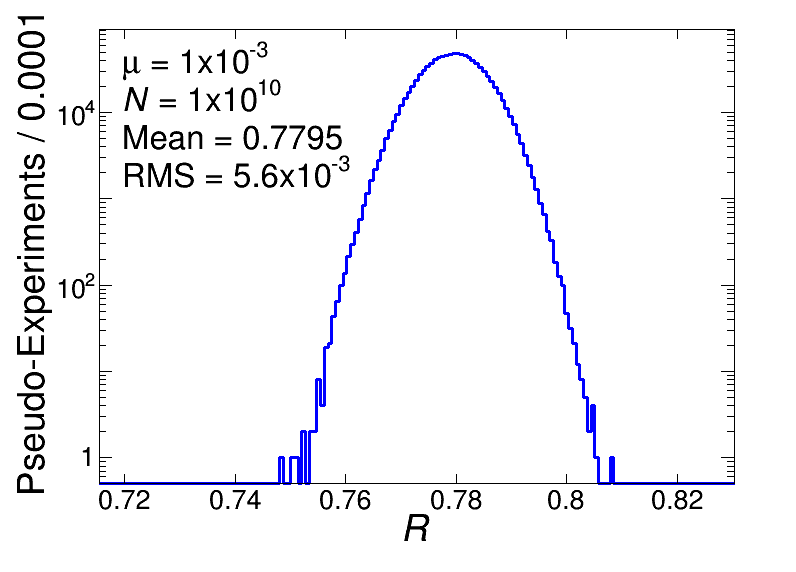}
	\label{subfig:ratio_muPtE-3_ev1e+10}
}
\caption{Distributions of \(R\) with \(N_{\text{PE}}=10^{6}\), for various small values of \(\mu\). In each case we have selected \(N\) large enough such that we are in the high-statistics regime to ensure a reliable estimation of \(R\), and we see that \(R\) converges to \(0.7795\) in all cases with small uncertainty.}
\label{fig:ratio_converging}
\end{figure}

A second conclusion is shown in Fig.~\ref{fig:ratio_converging}, which shows distributions of \(R\) measured in the high-statistics regime for various values of \(\mu\). We see that \(R\) converges to a constant number for small \(\mu\), i.e.\ for this particular differential cross section model and visible \(x\)-range it is \(R=0.7795\).

\section{\label{sec:closedStatSoln}Closed Form Statistical Study}

In this section we use a closed form calculation to study the simulation size needed to get reliable estimates of the constant multiplicative term. As previously noted, as \(N\) decreases, the \(R\) distribution transitions from being Gaussian to approximating a Cauchy distribution. To explain this, we note that the Cauchy distribution is the distribution of the ratio of two Gaussian random variables when the mean of the denominator is zero. When the mean of the Gaussian in the denominator is far enough away from zero, the distribution is Gaussian, and in the limit that it approaches zero, the distribution approaches the Cauchy distribution. Therefore, since both \(A^{\text{visible}}\) and \(A^{\text{inclusive}}\) have approximately Gaussian distributions, the distribution of \(R\) begins approximating a Cauchy distribution as \(\mu\) approaches zero. Cauchy distributions have a peak and a width, but the mean and standard deviation are undefined \cite{Cauchy}, and, without special considerations, determining these values using numerical methods is precarious and can give wrong values.

\begin{figure}[htbp]
\subfigure[]{
\centering
	\includegraphics[width=0.47\textwidth, keepaspectratio]{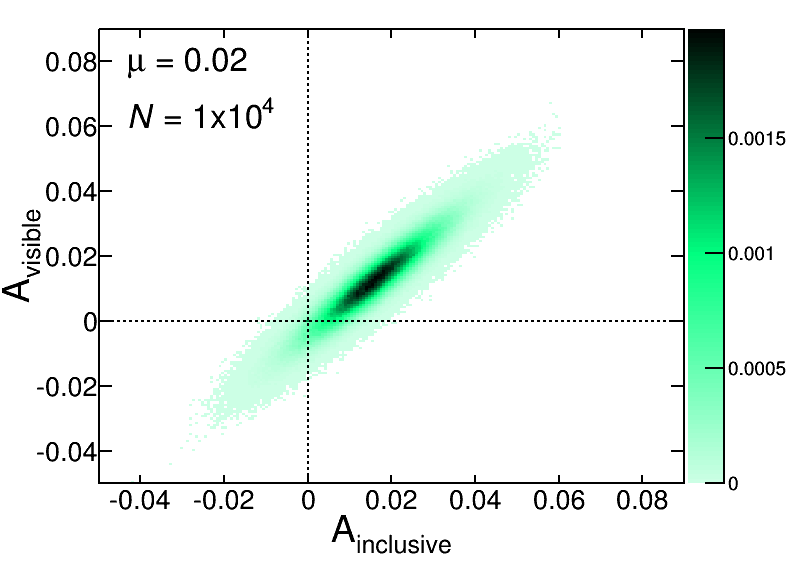}
	\label{subfig:tot-v-red-ev4}
}
\subfigure[]{
\centering
	\includegraphics[width=0.47\textwidth, keepaspectratio]{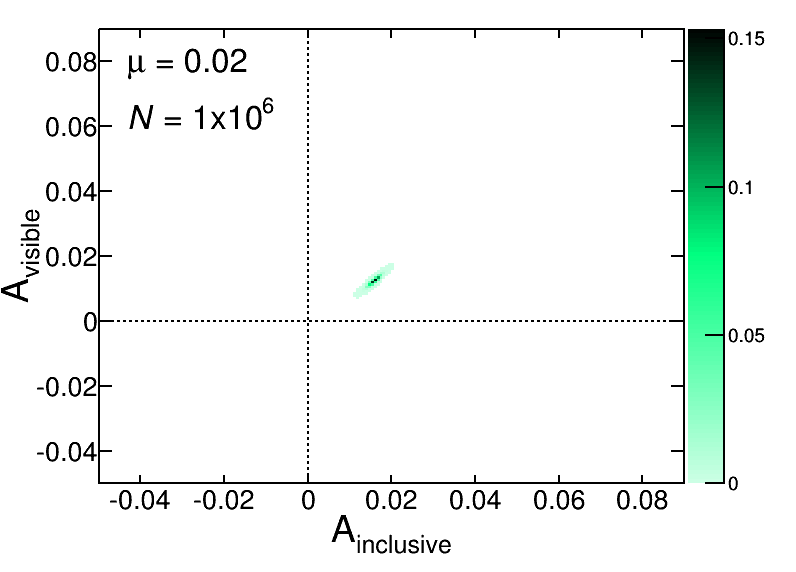}
	\label{subfig:tot-v-red-ev6}
}
\caption{Contour plots of \(A^{\text{visible}}\) vs. \(A^{\text{inclusive}}\) for \(N_{\text{PE}}=10^{6}\) and \(\mu=0.02\). We have set \(N=10^{4}\) and \(N=10^{6}\) in (a) and (b) respectively.}
\label{fig:tot-v-red-contours}
\end{figure}

In Fig.~\ref{fig:tot-v-red-contours} we show contour plots of \(A^{\text{visible}}\) vs. \(A^{\text{inclusive}}\) in both the low-statistics and high-statistics regimes, where we have taken \(\mu=2\times10^{-2}\) and \(N_{\text{PE}}=10^{6}\), with \(N=10^{4}\) and \(N=10^{6}\). We can think of our measurement of \(R\) as \(R=A^{\text{visible}}/A^{\text{inclusive}}=\tan(\theta)\), where \(\theta\) is the angle from the x-axis to the point on the plot measured from the origin. We can see that in the high-statistics regime, \(\theta\) doesn't vary much and is measuring the true slope of \(A^{\text{visible}}\) vs. \(A^{\text{inclusive}}\). However, in the low-statistics regime, \(\theta\) takes on all possible angles, and for a majority of the measurements \(\theta\) does not give a good measurement of the slope of \(A^{\text{visible}}\) vs \(A^{\text{inclusive}}\). This gives a visual demonstration of how the estimation of \(R\) breaks down below \(N_{\text{thresh}}\).

MC methods will not reliably estimate \(R\) if the measurement of \(A^{\text{inclusive}}\) (the denominator of \(R\)) from a single PE has a reasonable probability of being close to zero. Therefore, we need the simulation sample to be large enough such that \(A^{\text{inclusive}}\) is well separated from zero. We require \(A^{\text{inclusive}}\) to be at least \(k\) standard deviations away from zero, where \(k\) is typically a few, and calculate the minimum value of \(N\) that satisfies this constraint.

To do this we start with the inequality
\begin{linenomath}\begin{align}
A^{\text{inclusive}}&\geq k\sigma_{A^{\text{inclusive}}}\label{eq:afb3sigma}
\end{align}\end{linenomath}
where \(\sigma_{A^{\text{inclusive}}}\) is the uncertainty of the measured value of \(A^{\text{inclusive}}\). Using standard error propagation techniques, we calculate \(\sigma_{A^{\text{inclusive}}}\) to be
\begin{linenomath}\begin{align}
\sigma_{A^{\text{inclusive}}}&=\sqrt{\frac{1-{(A^\text{inclusive})}^2}{N}}.
\label{eqn:sigmaainc}
\end{align}\end{linenomath}
By plugging Eq.~(\ref{eqn:sigmaainc}) into Eq.~(\ref{eq:afb3sigma}) and solving for \(N\), we obtain a lower bound on \(N\) given \(A^\text{inclusive}\) that defines the high-statistics regime:
\begin{linenomath}\begin{align}
N&\geq\frac{k^2\Big(1-{(A^\text{inclusive})}^2\Big)}{{(A^\text{inclusive})}^{2}}\label{eq:mathematica_solution}.
\end{align}\end{linenomath}
In the limit where \(A\rightarrow0\), we find \(N\propto1/{(A^\text{inclusive})}^2\), which is consistent with what others have observed~\cite{rosner}. Since \(A^\text{inclusive}\propto\mu\), this is equivalent to \(N\propto1/\mu^2\), and when \(k\approx2\) we get a line that is consistent with the line shown in Fig.~\ref{fig:threshold}; any larger value of \(k\) will also give a good description of the high-statistics regime. While it is well known from similar calculations that a measurement of the uncertainty requires more statistics as \(A\) gets smaller, it is not as readily known just how important this is for use in MC correction techniques.

\section{\label{sec:conclusion}Conclusions}
We have studied the use of a simple multiplicative extrapolation method in asymmetry measurements. This method has already been used for measurements made at the Fermilab Tevatron of the \(t\bar{t}\) forward-backward asymmetry, and has potential for wide use. Perhaps most important for future experiments is that, if the correction factor and its uncertainty are to be estimated from a simulated sample, more statistics than expected may be needed, especially when the simulation yields small asymmetry values. We find that the number of simulated events needed for reliable measurements rises as \(1/{(A^\text{inclusive})}^{2}\).

\section*{Acknowledgements}
The authors would like to thank the Mitchell Institute for Fundamental Physics and Astronomy and the Department of Physics and Astronomy at Texas A\&M University, the DOE, and the Texas A\&M Office of Graduate and Professional Studies for their support. We would also like to thank Matteo Cremonesi, Ricardo Eusebi, Ulrich Husemann, Doug Orbaker, Jonathan Rosner, and Willis Sakumoto for their useful feedback.

\clearpage

\setcounter{figure}{0}    
\appendix
\section{\label{sec:closedNumSoln}Closed Form Numerical Validation}
\begin{figure}[htbp]
\centering
	\includegraphics[width=0.47\columnwidth, keepaspectratio]{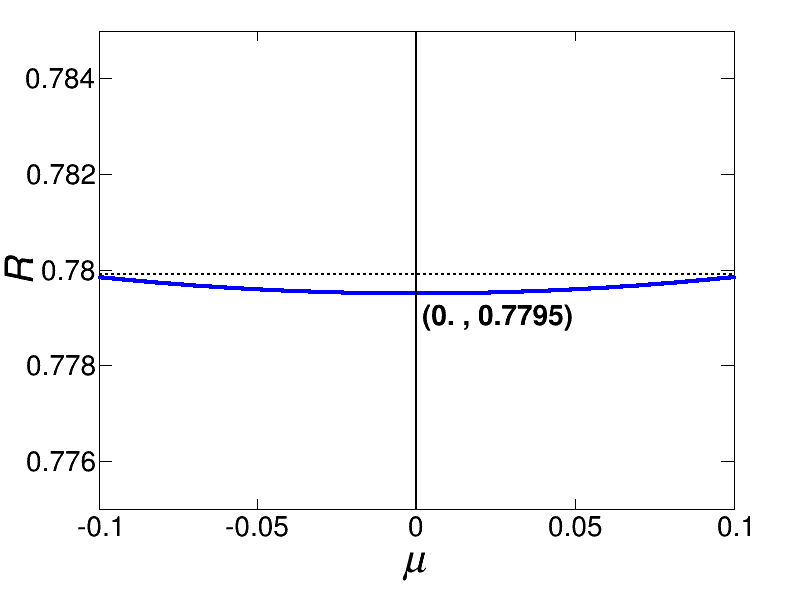}
\caption{A plot of \(R\) determined analytically as a function of \(\mu\). We can see here how in the limit of small \(\mu\), \(R=0.7795\) and only rises by \(0.04\%\) when \(\mu=0.1\).}
\label{fig:mathematica_ratioVSmu_plot}
\end{figure}

In this appendix we show both that the value of \(R\) is approximately constant for small values of \(\mu\) and that \(A^\text{inclusive}\) is linearly proportional to \(\mu\) for our single-Gaussian model~\cite{ziqing2014} using a closed form numerical solution. Specifically, we use the following equations,
\begin{linenomath}
\begin{align}
A^{\text{inclusive}}&=\frac{\bigintsss_{0}^{\infty}\text{dx}\big[\exp(-\frac{(x-\mu)^{2}}{2\sigma^{2}})-\exp(-\frac{(-x-\mu)^{2}}{2\sigma^{2}})\big]}{\bigintsss_{0}^{\infty}\text{dx}\big[\exp(-\frac{(x-\mu)^{2}}{2\sigma^{2}})+\exp(-\frac{(-x-\mu)^{2}}{2\sigma^{2}})\big]}\label{eqn:firstapp},\\
A^{\text{visible}}&=\frac{\bigintsss_{0}^{1.5}\text{dx}\big[\exp(-\frac{(x-\mu)^{2}}{2\sigma^{2}})-\exp(-\frac{(-x-\mu)^{2}}{2\sigma^{2}})\big]}{\bigintsss_{0}^{1.5}\text{dx}\big[\exp(-\frac{(x-\mu)^{2}}{2\sigma^{2}})+\exp(-\frac{(-x-\mu)^{2}}{2\sigma^{2}})\big]},\text{ and}\\
R&=\frac{A^{\text{visible}}}{A^{\text{inclusive}}}\label{eqn:lastapp},
\end{align}
\end{linenomath}
with \(\sigma=1.0\), to plot \(R\) as a function of \(\mu\), as \(\mu\) goes to 0. The result is shown in Fig.~\ref{fig:mathematica_ratioVSmu_plot}. While \(R\) is not exactly constant for all values of \(\mu\), it does not vary significantly from its value of 0.7795 at \(\mu=0\) in the regions that are typically relevant to experiments. For example, \(R\) only rises by \(0.04\%\) to 0.7798 at \(\mu=0.1\) (corresponding to \(A^{\text{inclusive}}=7.97\%\)). Similarly, \(R\) only rises by \(1.10\%\) at \(\mu=0.5\) (corresponding to \(A^{\text{inclusive}}=22.2\%\)). Thus, while assuming a constant multiplicative factor for the extrapolation is not perfect, the systematic uncertainty introduced from taking it to be constant is minimal for the region we are considering, and should be good for all but the highest precision measurements.

From Eq.~(\ref{eqn:firstapp}), we can see that this is the error function, thus we write that
\begin{linenomath}\begin{align}
A^{\text{inclusive}}&=\erf\Big(\frac{\mu}{\sqrt{2}}\Big),
\label{eqn:ainc_mu_rel}
\end{align}\end{linenomath}
where in the limit \(\mu\ll1\), \(\erf(\frac{\mu}{\sqrt2})\approx\sqrt{\frac{2}{\pi}}\,\mu\), and so we can see that \(A^\text{inclusive}\propto\mu\).

\clearpage

\bibliographystyle{elsarticle-num}
\bibliography{refs}{}

\begin{thebibliography}{10}
\expandafter\ifx\csname url\endcsname\relax
  \def\url#1{\texttt{#1}}\fi
\expandafter\ifx\csname urlprefix\endcsname\relax\def\urlprefix{URL }\fi
\expandafter\ifx\csname href\endcsname\relax
  \def\href#1#2{#2} \def\path#1{#1}\fi

\bibitem{costheta}
T.~Aaltonen, et~al., CDF Collaboration, Phys. Rev. Lett. 111 (2013) 182002.

\bibitem{CDFresult}
T.~Aaltonen, et~al., CDF Collaboration, Phys. Rev. Lett. 113 (2014) 042001.

\bibitem{d0Measurement2}
V.~M. Abazov, et~al., D0 Collaboration, Phys. Rev. D 88 (2013) 112002.

\bibitem{DZEROResult}
T.~Aaltonen, et~al., CDF Collaboration, Phys. Rev. D 89 (2014) 072001.

\bibitem{gausApprox}
T.~Aaltonen, et~al., CDF Collaboration, Phys. Rev. D 88 (2013) 072003.

\bibitem{cdfWillisAdditive}
T.~Aaltonen, et~al., CDF Collaboration, Phys. Rev. D 89 (2014) 072005.

\bibitem{d0willisversion}
V.~M. Abazov, et~al., D0 Collaboration, Phys. Rev. Lett. 115 (2015) 041801.

\bibitem{ziqing2014}
Z.~Hong, R.~Edgar, S.~Henry, D.~Toback, J.~S. Wilson, {and D. Amidei}, Phys.
  Rev. D 90 (2014) 014040.

\bibitem{ATLASAfb}
G.~Aad, et~al., {ATLAS Collaboration}, Euro. Phys. J. C 72 (2012) 1.

\bibitem{ATLASnote}
G.~Aad, et~al., ATLAS Collaboration, ATLAS-CONF-2015-048, PUBDB-2015-03920.

\bibitem{ATLAS2}
G.~Aad, et~al., ATLAS Collaboration, J. High Energy Phys. 02 (2014) 107.

\bibitem{ATLAS3}
G.~Aad, et~al., {ATLAS Collaboration}, J. High Energy Phys. 05 (2015) 061.

\bibitem{ATLAS4}
G.~Aad, et~al., ATLAS Collaboration, CERN-PH-EP-2015-217.

\bibitem{CMSAfb}
S.~Chatrchyan, et~al., CMS Collaboration, Phys. Lett. B 717 (2012) 129.

\bibitem{CMSsoph}
V.~Khachatryan, et~al., CMS Collaboration, CMS-TOP-13-013, CERN-PH-EP-2015-189.

\bibitem{CMS2014jua}
G.~Aad, et~al., ATLAS Collaboration, J. High Energy Phys. 02 (2014) 107.

\bibitem{LHCbBBAC}
R.~Aaij, et~al., LHCb Collaboration, Phys. Rev. Lett. 113 (2014) 082003.

\bibitem{CDFLJAfbl}
T.~Aaltonen, et~al., CDF Collaboration, Phys. Rev. D 88 (2013) 072003.

\bibitem{CDFLJAfb}
T.~Aaltonen, et~al., CDF Collaboration, Phys. Rev. D 87 (2013) 092002.

\bibitem{D0BmesonAfb}
V.~M. Abazov, et~al., D0 Collaboration, Phys. Rev. Lett. 114 (2015) 051803.

\bibitem{bbbar}
T.~Aaltonen, et~al., CDF Collaboration, Phys. Rev. D 92 (2015) 032006.

\bibitem{DZEROResult2}
V.~M. Abazov, et~al., D0 Collaboration, Phys. Rev. D 90 (2014) 072011.

\bibitem{Cauchy}
A.~Papoulis, ``Probability, Random Variables, and Stochastic Processes", 2nd
  ed., New York: McGraw-Hill, 1984.

\bibitem{rosner}
G.~Eilam, M.~Gronau, {and J. L. Rosner}, Phys. Rev. D 39 (1989) 819.

\end{thebibliography}
\end{document}